\documentclass[a4paper,10pt,twoside]{cpc-hepnp}

\usepackage{multicol}
\usepackage{graphicx}
\usepackage{booktabs}
\usepackage{amssymb,bm,mathrsfs,bbm,amscd}
\usepackage[tbtags]{amsmath}
\usepackage{lastpage}
\usepackage{CJK}

\begin{document}


\fancyhead[r]{Submitted to 'Chinese Physics C'}


\title{Analyzing the effects of post couplers in DTL tuning by the equivalent circuit model\thanks{Supported by National Natural Science Foundation of China (Major Research Plan Grant No. 91126003)}}

\author{%
      JIA Xiao-Yu$^{1,2}$
\quad ZHENG Shu-Xin$^{1,2;1)}$\email{zhengsx8@gmail.com}%
}
\maketitle

\address{%
$^1$ Key Laboratory of Particle $\&$  Radiation Imaging (TsinghuaUniversity),\  Ministry of Education,  Beijing,  100084, China\\
$^2$ Department of Engineering Physics, Tsinghua University, Beijing, 100084, China\\
}

\begin{abstract}
Stabilization of the accelerating field in Drift Tube Linac(DTL) is obtained by inserting Post Couplers(PCs). 
On the basis of the circuit model equivalent for the DTL with and without asymmetrical PCs, stabilization is deduced quantitatively: 
let $\delta \omega / \omega_0$ be the relative frequency error, then we discover that the sensitivity of field to perturbation is proportional to $\sqrt{\delta \omega / \omega_0}$ without PCs and to  $\delta \omega / \omega_0$ with PCs. 
Then we adapt the circuit model of symmetrical PCs for the case of asymmetrical PCs. 
The circuit model shows how the slope of field distribution is changed by rotating the asymmetrical PCs and  illustrates that the asymmetrical PCs have the same effect as the symmetrical ones in stabilization. 
\end{abstract}



\footnotetext[0]{\hspace*{-3mm}\raisebox{0.3ex}{$\scriptstyle\copyright$}2013
Chinese Physical Society and the Institute of High Energy Physics
of the Chinese Academy of Sciences and the Institute
of Modern Physics of the Chinese Academy of Sciences and IOP Publishing Ltd}%

\begin{multicols}{2}


\section{Introduction}

DTL is a kind of standing-wave structure and operates at TM010 mode. 
In this mode, the slope of dispersion curve is zero, which means the field distribution of this structure is sensitive to perturbations, such as beam loading, machining and installation errors.\cite{wangler2008rf,naito2003tuning,billen19853} 
We can reduce the effect of the perturbations by inserting the PCs, which are used to introduce resonant couple mode and increase the slop of dispersion curve at the operating point.\cite{knapp1970method}
 The tuning for DTL consists of the following tasks: resonant frequency of the structure, stabilization and distribution of accelerating electric field. 
Generally speaking, we can adjust slag tuners to get resonant frequency, insert the PCs into the right length to improve the stability of the field and rotate the PCs to adjust the field distribution precisely.\cite{machida1985stabilizing}

In order to reduce the interaction, the adjacent PCs are on different sides of drift tube and all PCs are perpendicular to stem. Both symmetrical and asymmetrical PCs can make the field stable. 
During the stabilization state, field distribution is flat when using symmetrical PCs, but we can get a tilt filed distribution with asymmetrical ones and the rate of slop can be changed by rotating the PCs.\cite{roy2007electromagnetic} 
Figure \ref{fig:DTL} shows the structure of DTL and the asymmetrical PCs.

References\cite{grespan2010circuital,vretenar2012low,ke1994continuous} propose an equivalent circuit model for the DTL with the symmetrical PCs, further, this article improves the model so that field distribution and stabilization with asymmetrical PCs will be analyzed and the rate of the tilting field can be estimated. 

\begin{center}
\includegraphics[width=6cm]{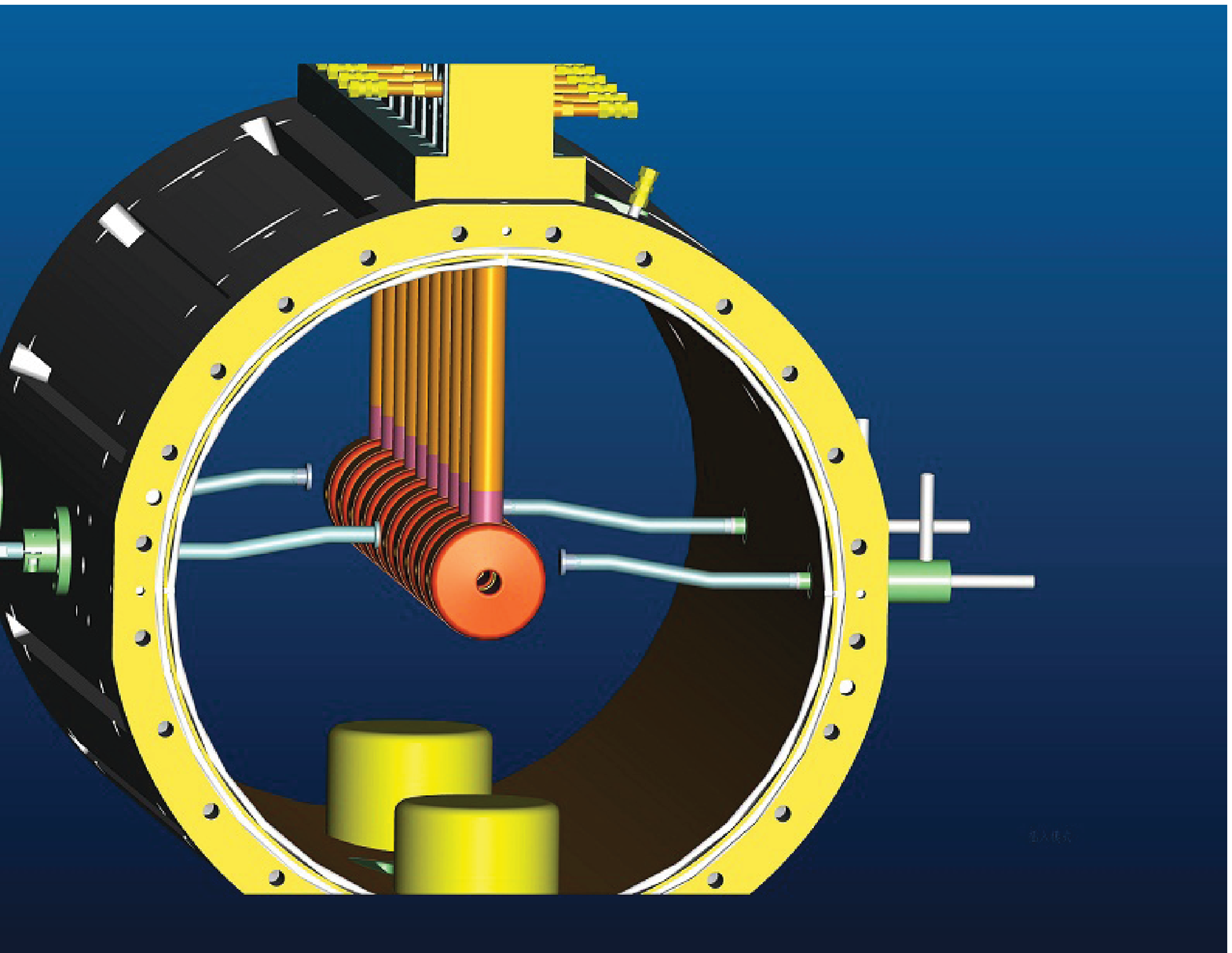}
\figcaption{\label{fig:DTL}   DTL and the post couplers(PCs).}
\end{center}

\section{The equivalent circuit for general periodic structure}

\begin{center} 
	\includegraphics[width=7cm]{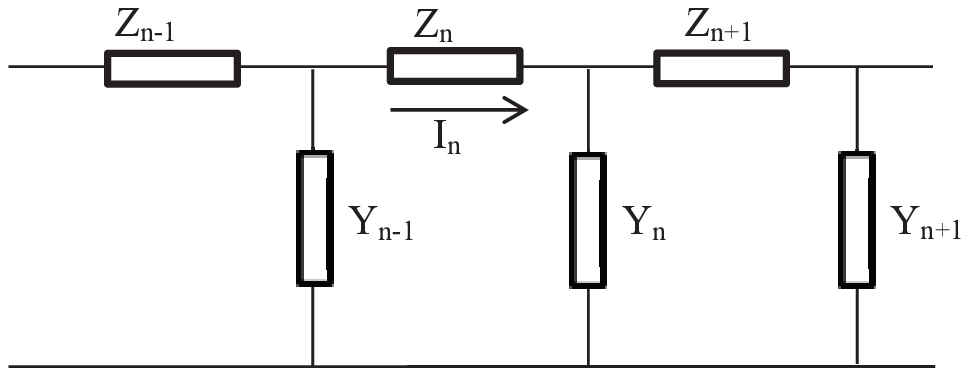}
	\figcaption{\label{fig:Cir-princ} The equivalent circuit for DTL without PCs. }   
\end{center}

General periodic structure can be represented by the circuit in Figure \ref{fig:Cir-princ}, which is composed of equivalent parallel admittance $Y_n$ and series impedance $Z_n$. By selecting the loop with $Y_{n-1}$, $Z_n$ and $Y_n$, we can get the recurrence formula of current $I_n$.
\begin{eqnarray} %
\label{equ:Circuit}
	-(I_{n-1}-I_n)/Y_{n-1} + I_n Z_n + (I_n - I_{n+1})/Y_n = 0.
\end{eqnarray}

The character of the periodic structure means $Y_{n-1}=Y_n=Y$ , $Z_n=Z$, then the recurrence formula can be simplified.
\begin{eqnarray} %
\label{equ:CircuitP}
	\left \{
		\begin{aligned}
      		&I_{n+1} - (2+YZ)I_n + I_{n-1} = 0    \qquad   &Y  \ne 0.\\
      		&I_{n+1}=I_n=I_{n-1}						   &Y= 0.\\
      	\end{aligned}
    \right .
\end{eqnarray}

$Y$ can not be equal to 0 in Equation (\ref{equ:Circuit}), and the situation of $Y=0$ is considered in Equation (\ref{equ:CircuitP}). There are different solutions for other situations of $YZ$.

1) $Y=0$: from Equation (\ref{equ:CircuitP}) we know that $I_{n+1}=I_n=I_{n-1}$. It is open circuit state and current through every cell is equal to each other. We call this situation the platmode.

2) $Z = 0$ and $Y \ne 0$: the solution is $I_n = I_{n-1} + \Delta I$. Current to next cell will be added by the same $\Delta I$, which is the current through admittance Y. Now the current distribution is an oblique line and we call this situation the tiltmode.

3) $YZ \ne 0$: the solution is 
\begin{eqnarray} %
\label{equ:solutionG}
	I_n=I_{n-1}e^{\pm \phi} \text{,} \cos\phi=1+{YZ}/{2}.
\end{eqnarray} %
YZ can be complex and $\phi$ also can be complex. This is the most general solution, but in an ideal state in DTL, power loss is not considered, and $Y$,$Z$ are near to 0 (that will be explained in Chapter 3). Then we will concentrate on several situations closely related to tuning.

4) $|YZ+2|<2$ and $YZ$ is real: the solution is
\begin{eqnarray} %
\label{equ:solutionT}
	I_n=I_{n-1}e^{\pm \phi}\text{,}\cos\phi=1+{YZ}/{2}.
\end{eqnarray} %
This is a special situation of Case 3). Here $\phi$ is real, and it is the phase difference between adjacent cells. This solution represents the transmission electromagnetic wave and $\phi$ is the phase shift between adjacent cells. We call this situation the transmissionmode.

5) $|YZ+2|>2$ and $YZ$ is real:the solution is
\begin{eqnarray} %
\label{equ:solutionC}
	I_n = I_{n-1}r \text{,} r=\frac{2+YZ \pm \sqrt{(2+YZ)^2-4}}{2}.
\end{eqnarray} %
It is also  a special situation exception of Case 3). The amplitude of current attenuates according to the exponential rate, $r$ is the current radio of adjacent cell. In DTL, this sulution means the electromagnetic wave cuts off, and the amplitude of field attenuates according to the exponential rate. But if $r$ is close to 1, the field distribution would be similar to an oblique line within suitable range.We call this situation the cutoff-mode.

 Table  \ref{tab:solutions} summarizes the important solutions. 

\end{multicols}
\begin{center}
\tabcaption{ \label{tab:solutions}  Solutions for Equation(\ref{equ:CircuitP}).}
\footnotesize
\begin{tabular*}{120mm}{@{\extracolsep{\fill}}ccc}
\toprule
Condition & Name for solution & Current relation  \\
\hline
$Z = 0$ and $Y \ne 0$&
	tiltmode  &
	$I_n = I_{n-1} + \Delta I$ \\

$Y = 0$ & 
	platmode & 
	$I_n = I_{n-1}$\\

$|YZ+2|<2$ and $YZ$ is real &
	transmissionmode &
	$I_n=I_{n-1}e^{\pm \phi}$,$\cos\phi=1+\frac{YZ}{2}$ \\

$|YZ+2|>2$ and $YZ$ is real&
	cutoffmode	&
	$I_n = I_{n-1}r ,r=\frac{2+YZ \pm \sqrt{(2+YZ)^2-4}}{2}$\\
\bottomrule
\end{tabular*}%
\end{center}

\begin{multicols}{2}

As is known, Y and Z are as functions of the frequency $\omega$, so different frequncies correspond to different modes.
If the RF source frequency differs from the structure resonant frequency by  $\delta \omega$, then the field distribution would deviate from the design value. 
For the cutoffmode, stabilization can be measured by the r, which is the current radio of adjacent cell, and for the transmissionmode, it can be measured by the phase change of $\phi$. 
Then we will prove this change is proportional to  $\delta \omega / \omega_0$ with PCs in the right position, and proportional to  $\sqrt{\delta \omega / \omega_0}$ without PCs. For example, the operating frequency is 325MHz, and the difference between the RF source frequency and the structure resonant frequency is less than 30kHz. In this case, the field distribution would change by about 0.01\% with PCs and about 1\% without PCs. We can see that PC can improve the stabilization significantly.


\section{The equivalent circuit for DTL}

DTL is a kind of quasi-periodic structure, but in this paper we will approximately regard it as a periodic one. The power loss is neglected, so there is no resistance in circuit.

\subsection{The equivalent circuit for DTL without PCs}

\begin{center} 
	\includegraphics[width=7cm]{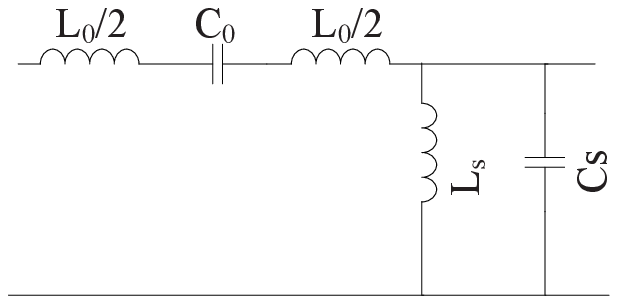}
	\figcaption{\label{fig:Cir-basic} The equivalent circuit for DTL without PCs.}
\end{center}

In the circuit model in Figure \ref{fig:Cir-basic}, $C_0$ is the capacitance between two drift tubes; $L_0$ is the inductance of the drift tube; $C_s$ is the capacitance between drift tube and tank; $L_s$ is the inductance of the stem. There are two resonant units in the circuit: one consists of $L_0$ and $C_0$ and the other is made up by $L_s$ and $C_s$. Define $\omega_0=1/\sqrt{ L_0 C_0 }$   and $\omega_s=1/\sqrt{L_s C_s }$  . As we know, $\omega_s$ is much lower than $\omega_0$, the operating frequency. So we can regard $L_s$ and $C_s$ as a capacitance $C$ parallel in the circuit.\cite{grespan2010circuital}
\begin{eqnarray}
\label{equ:DefineC}
       C = C_s \frac{\omega^2 - \omega_s^2}{\omega^2}.
\end{eqnarray}

Voltage of the capacitance $C_0$ can be considered as the product of cell length and average accelerating field:
$u_n = L_{cell} E_{n,ave}$
, so the current $i_n$ through $C_0$ is related to the field:
$i_n=j \omega C_0 u_n= j \omega C_0 L_{cell} E_{n,ave}$.

The operating frequency is equal to the resonant frequency of DTL, which means $\omega=\omega_0$, so $Z=0$, $Y=j\omega_0 C$. 
According to Table 1, it is the tiltmode and the slope of field is decided by boundary condition, which is the end cavities frequency of DTL. 
In practice, we can tilt the field by changing the frequency of the end cavities. 
When the operating frequency is shifted by $ \delta \omega $, by linear approximation we can get
\begin{eqnarray}
\label{equ:YZ1}
	YZ = \frac{C}{C_0} \frac{\omega_0^2 - (\omega_0 + \delta \omega)^2}{\omega^2}
	\approx  -2  \frac{C}{C_0}  \frac{\delta \omega}{\omega_0}.
\end{eqnarray}

For $\delta \omega / \omega \ll 1$, so $|YZ| \ll 1$, 
if $\delta \omega > 0$, then $ YZ < 0 $, 
according to Table 1, the field would stay in transmissionmode and 
$\Delta \phi \approx = \sqrt{2C \delta \omega /(C_0 \omega_0)}$ ;
$ \delta \omega > 0 $, the field is in cutoffmode and 
$\Delta r \approx \pm \sqrt{2C \delta \omega /(C_0 \omega_0)}$. 
Because $C/C_0$  is constant, the sensitivity of field to perturbation is in proportion to $\sqrt{\delta \omega / \omega_0}$, and the stabilization is poor.

\subsection{The equivalent circuit for DTL with symmetrical PCs}

\begin{center} 
	\includegraphics[width=8cm]{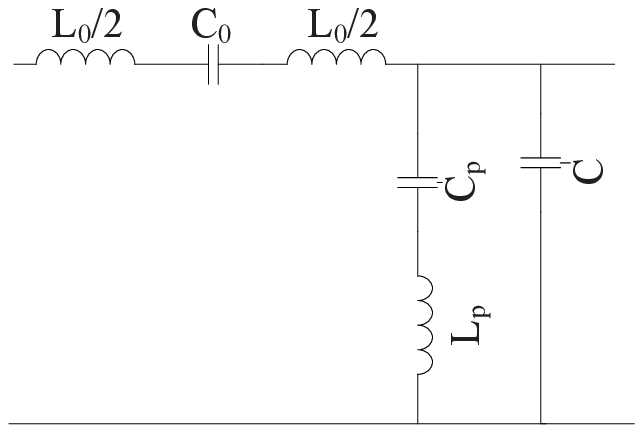}
	\figcaption{\label{fig:Cir-3} The equivalent circuit for DTL without PCs. }
	
\end{center}

$C_p$ in Figure \ref{fig:Cir-3} is the capacitance between PC and drift tube and $L_p$ is the inductance of PC, then define $\omega_p=1/\sqrt{L_p C_p} $. 
By inserting PCs, we introduce a new resonant mode, and the stabilization can be got if this mode is completely coupled with the operaton mode. 
Coupling condition can be proposed by solving the dispersion eqution\cite{wangler2008rf}
\begin{eqnarray} %
\label{equ:disperse}
	\cos \phi = 1 + \frac{Y(\omega)Z(\omega)}{2}.
\end{eqnarray} 
It exactly means the parallel admittance Y=0. Corresponding to Table 1, it is the flatmode and the distribution of field is unrelated to boundary conditions.
\begin{eqnarray}
\label{equ:CouplingCondion}
	\omega_p^2 = \frac{C}{C+C_p}\omega_0^2.
\end{eqnarray}

When the difference between the RF source frequency and  the structure resonant frequency  is $\delta \omega$, by linear approximation we can get
\begin{eqnarray}
\label{equ:YZ2}
	YZ \approx -4\frac{C(C+C_p)}{C_0 C_p} \frac{\delta \omega^2}{\omega_0^2}.
\end{eqnarray}

$|YZ| \ll 1$ and $YZ < 0$, so the structure operates in the transmissionmode and 
\begin{eqnarray}
\Delta \phi \approx \frac{\delta \omega}{\omega} \sqrt{\frac{C(C+C_p)}{C_0 C_p}}. 
\end{eqnarray}
The sensitivity of field to perturbation is in proportion to $\delta \omega / \omega$, and the stabilization is improved compared with that without PCs.

\subsection{The equivalent circuit for DTL with asymmetrical PCs}
 
\begin{center}
	\includegraphics[width=7cm]{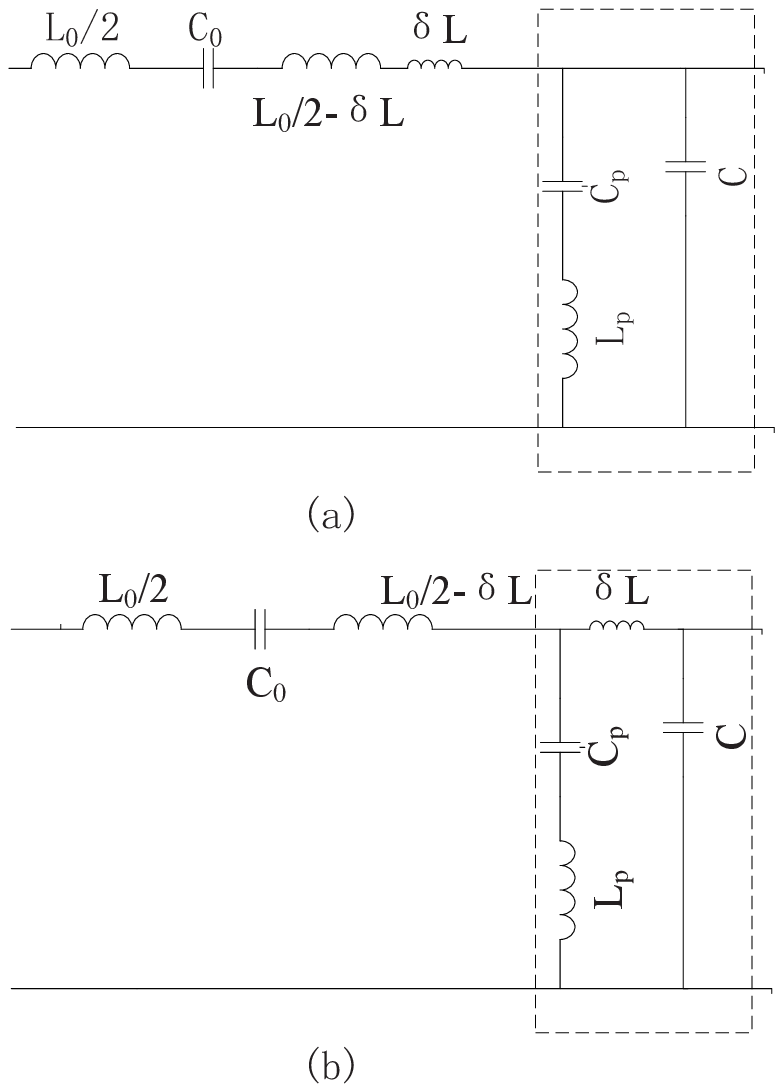}
	\figcaption{\label{fig:Cir4} The circuit for DTL with symmetrical PCs(a) and with asymmetrical PCs(b). }
\end{center}

\begin{center}
	\includegraphics[width=7cm]{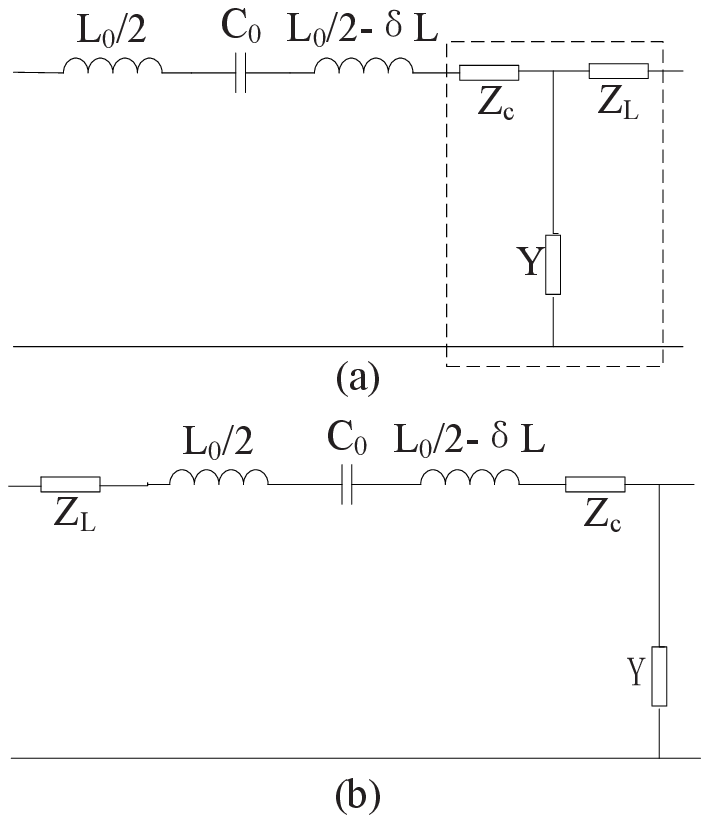}
	\figcaption{\label{fig:Cir5} The result  for  $\Delta-Y$ transform. }
\end{center}

While rotated, the tip position of asymmetrical PCs changes, so the positions of $C_p$ and $L_p$ in circuit change at the same time, as Figure \ref{fig:Cir4}(b). 
$\Delta-Y$ transform is taken for the box in Figure \ref{fig:Cir5}(b), so we can get a simple circuit with the same form of Figure \ref{fig:Cir-princ}. 
Because of  $\delta L/L_0 \ll 1$, when PCs are tuned to satisfy  Equation (\ref{equ:CouplingCondion}), we can get
\begin{eqnarray}
\label{equ:YZ3}
	YZ \approx 
		\left (
		\frac {C \delta L}{C_0 L_0}
		\right )^2.
\end{eqnarray}

The structure works in the cutoffmode, and
 $r \approx 1 \pm C \delta L/(C_0 L_0 )$. 
Because $r$ (defined in Chapter 2)is quite close to 1, which means the distribution of the current would look like a tilt line whose slope would be affected by the end cavities and the value of $\delta L$. 
Slope of the amplitude of electric field can be adjusted by varying the value of $\delta L$, that means rotating PCs.
Then we analyze the stabilization with the above method. 
If
\begin{eqnarray}
\label{equ:ConditionRotation}
	\frac{\delta \omega}{\omega} \ll 1,
	\frac{\delta L}{L_0} \ll 1,
	\frac{\delta \omega}{\omega} / \frac{\delta L}{L_0} \ll 1.
\end{eqnarray}
is satisfied, the result will be obvious. 
$\Delta r \approx \pm C \delta \omega /(C_0 \omega_0)$ , it illustrates that asymmetrical PCs have the same effect as the symmetrical ones in stabilization.

\section{Conclusion}
On the basis of the equivalent circuit model, the character of the stabilization and distribution of the electric field in DTL was analyzed in this article. 
Firstly, the stabilization was described by the relationship between field perturbation and frequency difference: 
the sensitivity of field to perturbation is in proportion to 
$\sqrt{\delta \omega / \omega_0}$ 
without PCs and to 
$\delta \omega / \omega_0$ 
with PCs . 
Secondly, the difference between the symmetrical and asymmetrical PCs is that the electric field distribution is flat and unrelated to the adjustment of the end cavities with the symmetrical PCs, but that attenuats according to exponential rate with asymmetrical PCs. 
If r is close to 1, the field distribution can be considered as an oblique line and the slope of it can be changed by rotating the asymmetrical PCs.

\end{multicols}

\vspace{10mm}
\vspace{-1mm}
\centerline{\rule{80mm}{0.1pt}}
\vspace{2mm}


\begin{multicols}{2}

\end{multicols}

\clearpage

\end{document}